\documentclass[reprint,
superscriptaddress,
nofootinbib,
aps,
pra,
showkeys
]{revtex4-2}
\pdfoutput=1

\usepackage{amsthm, mathtools, amssymb}
\usepackage{braket, physics} 
\usepackage{nicefrac}

\usepackage{orcidlink}
\graphicspath{ {./images/} }

\usepackage{graphicx, placeins, float}
\usepackage{multirow, makecell, hhline} % For tabular
\usepackage[caption=false]{subfig}
\usepackage[export]{adjustbox}

\usepackage{xcolor}

\usepackage{hyperref}
\usepackage[nameinlink,capitalize]{cleveref}
\hypersetup{
	colorlinks = true,
	linkbordercolor = {white},
	linkcolor={purple},
	citecolor={purple},
	urlcolor={blue}
}

\Crefname{appendix}{Supplementary Material}{Supplementary Materials}

\newcommand*\dif{\mathop{}\!\mathrm{d}}

\theoremstyle{definition}
\newtheorem{theorem}{Theorem}
\makeatletter
\def\thm@space@setup{\thm@preskip=0pt
\thm@postskip=0pt}
\makeatother
\newtheorem*{theorem*}{Theorem}
\makeatletter
\def\thm@space@setup{\thm@preskip=0pt
\thm@postskip=0pt}
\makeatother

\newcommand{\I}{\mathcal{I}}
\newcommand{\U}{\mathcal{U}}

\newcommand{\F}{F}
\renewcommand{\O}{\mathcal{O}}

\begin{document}
\title{Quantum optical shallow networks}
\author{Simone Roncallo\,\orcidlink{0000-0003-3506-9027}}
%\author{Simone Roncallo}
	\email[Simone Roncallo: ]{simone.roncallo01@ateneopv.it}
	\affiliation{Dipartimento di Fisica, Università degli Studi di Pavia, Via Agostino Bassi 6, I-27100, Pavia, Italy}
	\affiliation{INFN Sezione di Pavia, Via Agostino Bassi 6, I-27100, Pavia, Italy}
	
\author{Angela Rosy Morgillo\,\orcidlink{0009-0006-6142-0692}}
%\author{Angela Rosy Morgillo}
	\email[Angela Rosy Morgillo: ]{angelarosy.morgillo01@ateneopv.it}
	\affiliation{Dipartimento di Fisica, Università degli Studi di Pavia, Via Agostino Bassi 6, I-27100, Pavia, Italy}
	\affiliation{INFN Sezione di Pavia, Via Agostino Bassi 6, I-27100, Pavia, Italy}

\author{Seth Lloyd\,\orcidlink{0000-0003-0353-4529}}
%\author{Seth Lloyd}
	\email[Seth Lloyd: ]{slloyd@mit.edu}
	\affiliation{Massachusetts Institute of Technology, Cambridge, MA 02139, USA}
	
\author{Chiara Macchiavello\,\orcidlink{0000-0002-2955-8759}}
%\author{Chiara Macchiavello}
	\email[Chiara Macchiavello: ]{chiara.macchiavello@unipv.it}
	\affiliation{Dipartimento di Fisica, Università degli Studi di Pavia, Via Agostino Bassi 6, I-27100, Pavia, Italy}
	\affiliation{INFN Sezione di Pavia, Via Agostino Bassi 6, I-27100, Pavia, Italy}
	
\author{Lorenzo Maccone\,\orcidlink{0000-0002-6729-5312}}
%\author{Lorenzo Maccone}
	\email[Lorenzo Maccone: ]{lorenzo.maccone@unipv.it}
	\affiliation{Dipartimento di Fisica, Università degli Studi di Pavia, Via Agostino Bassi 6, I-27100, Pavia, Italy}
	\affiliation{INFN Sezione di Pavia, Via Agostino Bassi 6, I-27100, Pavia, Italy}
	
\begin{abstract}
    Classical shallow networks are universal approximators. Given a sufficient number of neurons, they can reproduce any continuous function to arbitrary precision, with a resource cost that scales linearly in both the input size and the number of trainable parameters. In this work, we present a quantum optical protocol that implements a shallow network with an arbitrary number of neurons. Both the input data and the parameters are encoded into single-photon states. Leveraging the Hong-Ou-Mandel effect, the network output is determined by the coincidence rates measured when the photons interfere at a beam splitter, with multiple neurons prepared as a mixture of single-photon states. Remarkably, once trained, our model requires constant optical resources regardless of the number of input features and neurons.
\end{abstract}
\keywords{Quantum classifier; Quantum optical neuron; Quantum neural networks; Hong-Ou-Mandel effect;}
\maketitle

\section{INTRODUCTION}
Machine learning is often framed as the task of approximating an unknown function: a continuous mapping in regression, or a decision boundary in classification (with classes at least partially separable in the feature space). Classical shallow neural networks, i.e. networks with one hidden layer of $M$ neurons, are known to be universal function approximators~\citep{book:Goodfellow}. In the uniform norm, they can approximate any continuous function arbitrarily well (on compact subsets of its domain) as 
\begin{equation}
    f_{wW}(x) = \textstyle\sum_{i=0}^{M-1}w_ig(W_i\cdot x + \beta_i ) \ \text{with} \ x \in \mathbb{R}^N
    \label{eq:UniversalApproximator} 
\end{equation}
and $M \in \mathbb{N}$, $\{W_i \in \mathbb{R}^N : i=0,\ldots,M-1\}$ (hidden parameters), $w,\beta \in \mathbb{R}^M$ (output parameters and biases) and $g$ a non-constant, bounded, and continuous activation function. Different formulations of this result have been proposed, generalizing it to several cases such as unbounded activation functions~\citep{cybenko1989approximation,hornik1989multilayer,hornik1991approximation,
mhaskar1992approximation,leshno1993multilayer,barron1993universal,chen1995universal}. Despite several attempts to estimate $M$ a priori, achieving the desired expressivity can be a resource-hungry task: a shallow network requires $\mathcal{O}(MN)$ computational operations.

A fundamental routine in computer vision is image classification~\citep{lecun1998gradient,krizhevsky2017imagenet,he2016deep,dosovitskiy2021image}, which has applications from medical diagnosis~\citep{cai2020review,dao2024recent} to security and social media analysis~\citep{babu2021survey,long2024crisisvit,amerini2019social}. In this case, the cost of classifying the image adds to that of optically acquiring it. Conventional imaging protocols require a number of photons that scales linearly in the image resolution, i.e.~$\mathcal{O}(N)$: one photon per pixel to reconstruct a black-and-white natural image.

Research has recently focused on finding alternative implementations of neural networks, exploiting quantum mechanical effects in computational methods~\citep{lloyd2013quantum,cai2015entanglement,tacchino2019artificial,benatti2019continuous,mangini2020quantum} or optical implementations~\citep{steinbrecher2019quantum,killoran2019continuous,sui2020review,zhang2021quantum,spall2022hybrid,stanev2023deterministic,wood2024kerr,spall2025training,hoch2025quantum}, with applications including image classification~\citep{slabbert2025classical, sun2025scalable,sakurai2025quantum}.
\begin{figure}[b]
	\centering
	\includegraphics[width = 0.45 \textwidth]{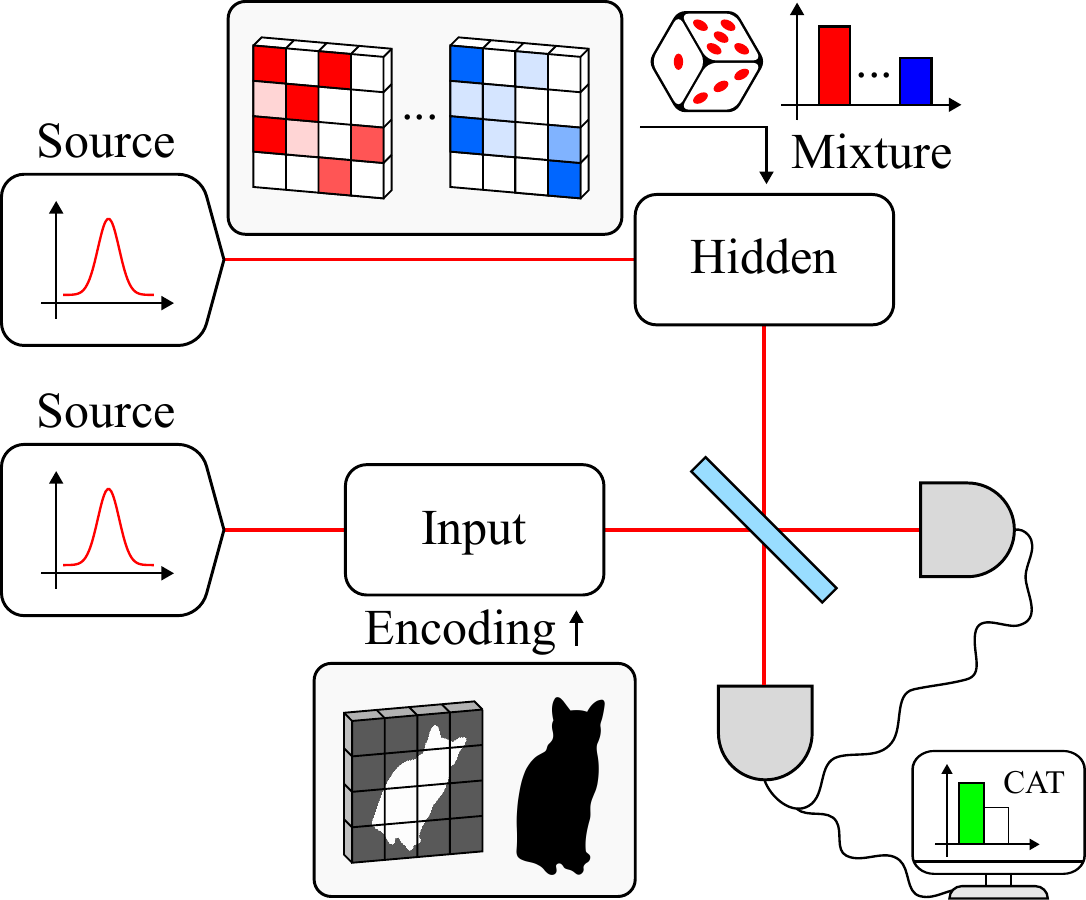}%
	\caption{\label{fig:Setup}Quantum optical shallow network implemented by a mixed state. One photon collects the input data, e.g. the shape of an object, the other encodes a sequence of hidden parameters, randomly sampled from a set of $M$ different configurations with probabilities acting as output parameters. The input is classified by two bucket detectors that measure the rate of coincidences at the beam splitter output.}
\end{figure}

In this paper, we present a quantum optical version of a classical shallow neural network. We consider the Hong-Ou-Mandel effect~\citep{hong1987measurement,brańczyk2017hong}, in which the probability that two photons exit a beam splitter in different modes, depends on their spectral distinguishability, with applications to kernel methods~\citep{bowie2023quantum} and optical perceptrons~\citep{roncallo2025quantum}. In our method, the spatial modes of a pure photon encode the input features, given by diffraction on a free-space object or as data encoded by a spatial light modulator (SLM). Then, a mixture of single-photon states encodes $M$ set of trainable parameters (the analogue of the network hidden layer). We show that the response function of this apparatus is that of a universal approximator with additional normalization constraints, yielding \cref{eq:UniversalApproximator} from the rate of photons coincidences at the output ports. See \cref{fig:Setup} for a representation of this architecture. We demonstrate that our protocol requires constant resources in inference, both in terms of photons and computational operations: a superexponential speedup over its classical counterpart. Finally, we numerically assess its approximation capabilities by verifying that increasing the number of neurons leads to improved accuracy in image classification tasks. Our framework provides a theoretical protocol adaptable to different experimental platforms. An actual implementation is under investigation~\citep{sciarrino2026quantum}.

\section{PROTOCOL}
In this section we design an optical apparatus whose response function is mathematically equivalent to that of a universal approximator according to \cref{eq:UniversalApproximator}. Consider two separable single-photon states, distinctively fed into the top and left input modes of a balanced beam splitter. In the left branch, labelled by mode $a$, a pure single-photon encodes the input state to be processed and classified, e.g.~the spectral amplitude reflected off by an unknown free-space object, or a pattern encoded by a spatial light modulator. In the top branch, labelled by mode $b$, a single-photon density operator encodes the parameters of the hidden layer, namely $M$ independent sets, each corresponding to a single hidden neuron of $N$ trainable (user-controlled) parameters. We call $a$ and $b$ input and hidden modes, respectively. After the beam splitter, we consider two bucket detectors that click whenever a photon exits their mode, without retrieving any spatial information. Due to the Hong-Ou-Mandel effect, the coincidence probability at the output, i.e. the probability that both detectors click, depends on the spectral distinguishability between the input and the hidden modes. When the hidden state is fixed, we show that measuring the rate of coincidences is equivalent to computing the output of the classical shallow network of \cref{eq:UniversalApproximator}. We report a schematic representation of this apparatus in \cref{fig:Setup}

In the input branch, consider a pure multimode single-photon state, namely
\begin{equation}
	\ket{\I} = \int \dif^2k \ \I(k) a^\dagger_\omega(k) \ket{0} \ .
	\label{eq:InputState}
\end{equation}
The input $\I(k)$ is encoded in the photon spatial modes, with $a_\omega^{\dagger}(k)$ the creation operator, $\ket{0}$ the vacuum state, and $k$ the two-dimensional momentum conjugated to the transverse coordinates on the image plane. \cref{eq:InputState} is derived by considering a monochromatic source with frequency $\omega$. The source emits a single photon towards the encoding system, e.g. a free-space object or a SLM. Hence, $\I$ contains the spectral information of the source and of the input, as well as the transfer functions of the free-space propagation and of an eventual linear optical apparatus.

In the hidden branch, consider a mixed single-photon state, namely a mixture of $M$ pure single-photon components $\{\ket{W_{\lambda_{i}}} : i = 0,\ldots,M-1\}$ with probability $w_i\geq 0$, each controlled by a set of $N$ trainable parameters, such that $\{\lambda_{i}\in\mathbb{R}^N : i = 0, \ldots , M-1 \} \to \ket{W_{\lambda_{i}}}$. This state is described by the density operator $\rho_{\U_{\lambda}} = \sum_i w_i \ket{W_{\lambda_i}}\!\bra{W_{\lambda_i}}$, that is
\begin{gather}
    \rho_{\U_\lambda} = \int \dif^2k_1 \dif^2k_2 \ \U_\lambda(k_1,k_2)b_\omega^{\dagger}(k_1)\ket{0}\!\bra{0}b_\omega(k_2) \ , \label{eq:MixedState} \\
	\text{with} \ \U(k_1,k_2) = \sum_{i=0}^{M-1} w_i W_{\lambda_i}(k_1)W^*_{\lambda_i}(k_2) \ ,
\end{gather}
subject to the normalization conditions $\sum_{i} w_i = 1$ and $\|W_{\lambda_i}\|^2 = 1 \ \forall i$, with $\|\cdot\|$ denoting the $L^2$-norm. This state can be generated directly or indirectly: by subsequently feeding the hidden branch with pure states agnostically sampled from $\rho_{\U_\lambda}$, with probability $w_i$, or by measuring the rate of coincidences of $\ket{W_{\lambda_i}}$ through $M$ distinguishable experimental runs, taking their weighted average over $w_i$. Although computationally more expensive, the latter strategy provides more freedom, bypassing the normalization and positivity constraints on $w_i$ (allowing the exploration of a larger subspace of parameters during training). At the end of the section, we discuss how different generation protocols affect the resource cost of a single classification instance.

Consider a bucket detector, namely a photodetector without spatial sensitivity, placed at the output of each branch. After feeding both states into the beam splitter, we describe coincidences in terms of the projector that finds a photon in the collective modes $a$ and $b$ integrated on $k$ (bucket detectors), i.e. $\Pi_a \otimes \Pi_b$. For a single pure component $\ket{W_{\lambda_i}}$, the probability that both detectors click reads $\text{Tr}[U_{H}\rho_iU^{\dagger}_{H} \Pi_a \otimes \Pi_b] = \left(1 -  |\langle \I , W_{\lambda_i} \rangle|^2 \right)/2$, where $U_{H}$ is the beam splitter unitary and $\rho_i$ the input-hidden bipartite density operator, corresponding to $\ket{\I}\otimes\ket{W_{\lambda_i}}$. For a mixture $\rho = \sum_i w_i \rho_i$, the coincidence probability reads $p(1_a\cap 1_b)=[1-f_{wW}(\I)]/2$, with
\begin{equation}
	f_{wW}(\I) = \sum_{i=0}^{M-1}w_i|\langle \I , W_{\lambda_i} \rangle|^2 \ ,
	\label{eq:Mixture}
\end{equation}
and $\langle \cdot , \cdot \rangle$ the $L^2$ inner product. Measurements apply a nonlinear map to such product: an essential step to approximate a neuron and overcome the limitations of linear regression. See \cref{app:models} for a complete derivation from Wick's theorem. We finalize our model by post-processing the output with a bias $\beta\in\mathbb{R}$, for example, to counteract unbalanced training datasets or mitigate the effect of photon loss, and a subsequent sigmoid activation function $\sigma(x) = 1/[1+\exp(-x)]$, that enhances the overall nonlinearity. Namely, the post-processed output reads $F_{\theta} = \sigma(f_{w W}(\I) + \beta)$, with $\theta$ denoting the set of trainable parameters $\{w, W_\lambda,\beta\}$. 
Importantly, both photons have the same frequency $\omega$, which can be always enforced by a proper choice of the longitudinal momentum.  Experimental mismatches can be solved by optimizing the hardware setup, or by introducing an additional trainable parameter to improve visibility~\citep{crum2025mode}.
\begin{figure}[t]
    \centering
    \includegraphics[trim={4 0 5 0},clip,width=.95\linewidth]{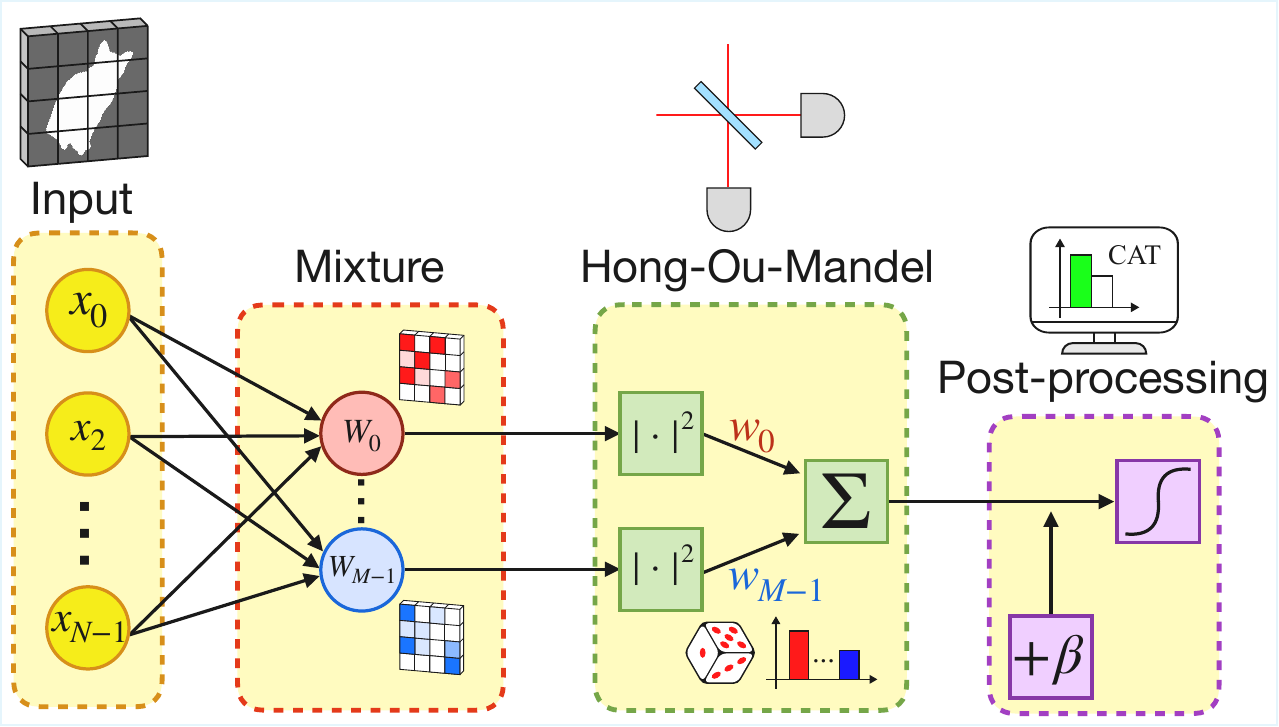}%
    \caption{\label{fig:Analogy}Analogy between a classical shallow neural network, i.e.~a network with $M$ hidden neurons, and our interferometric setup. The nodes $x_j$ are related to the input spectral amplitudes, encoded by the spatial light modulator on the left branch of the interferometer. Each pure component of the mixture encodes $N$ parameters of a different hidden neuron $W_i$, with square absolute value activation function and no biases (a total of $M$ sets of parameters). The output weights $w_i$, eventually normalized and positive, are related to the probabilities attached to each component of the mixture. Finally, a sigmoid activation function and a bias are introduced in post-processing.}
\end{figure} 

\cref{eq:Mixture} is precisely the response function of a classical shallow neural network.  To see this, let the input and the hidden states be implemented by two amplitude-modulated SLMs of $N$ pixels. In the left branch, $N$ input features $\{x_j\}$ (e.g.~the pixel values of a digital image) are encoded in the spatial spectrum $\hat{\I}(r) = \sum_{j}x_j \delta(r-r_j)$, with $r\in\mathbb{R}^2$ the transverse spatial coordinate on the image plane and $r_j$ the position of each pixel on the SLM, where $j=0,\dots,N-1$. The hat operator denotes the two-dimensional Fourier transform on such plane. In the hidden branch, each component of $ \rho_{\U_\lambda}$ reads $\hat{W}_{\lambda_{i}}(r) = \sum_{j}(W_{\lambda_i})_j \delta(r-r_j)$. By regularizing the Dirac delta as a sequence of approximating functions, we get $ f_{wW}(\I) = \sum_{i=0}^{M-1}\sum_{j=0}^{N-1}w_i| (W_{\lambda_i})_jx_j |^2$. Identifying $(W_{\lambda_i})_j \leftrightarrow W_{ij}$, we obtain \cref{eq:UniversalApproximator} for a square absolute value activation function with no biases: a universal approximator with additional constraints, due to the $L^2$ and $L^1$ normalization of the hidden and output layers, respectively. We report a schematic representation of this analogy in \cref{fig:Analogy}.

Classically, a shallow neural network (with $N$ input features and $M$ hidden neurons) has a computational cost that scales at least linearly in the number of mathematical operations. Indeed, a single classification instance requires the (subsequent or parallel) computation of $M$ scalar products of vectors of size $N$, namely $\mathcal{O}(MN)$ resources. When features are pixels values, an additional optical cost arises due to image acquisition. In the simplest scenario, i.e. digital black-and-white images, such cost scales linearly with the resolution, namely $\mathcal{O}(N)$ photons. Our method exploits the Hong-Ou-Mandel effect, performing such resource-hungry calculations at the optical level, i.e. with constant computational resources: $\mathcal{O}(1)$ mathematical operations only determined by the post-processing application of a single bias and a scalar sigmoid activation function~\citep{roncallo2025quantum}. Optically, the computation is not deterministic: we must take into account the number of statistical samples, i.e. photons, needed to estimate $f_{wW}(\I)$ in terms of the probability $p(1_a\cap1_b)$. For the $i$th pure state $\rho_i$ of a mixture $\rho = \sum_i w_i \rho_i$, estimating the rate of coincidences $\text{Tr}[U_{H}\rho_iU^{\dagger}_{H} \Pi_a \otimes \Pi_b]$ is equivalent to reconstructing a binomial distribution from the empirical frequencies with uncertainty $\varsigma$ (with success-failure corresponding to the presence-absence of such a coincidence). Such estimation requires $\mathcal{O}(\varsigma^{-2})$ photons. This is a constant overhead, independent of the number $N$ of input features encoded in $\rho_i$. Consider now the mixture of $M$ pure states. Reconstructing $p(1_a\cap1_b)$ by separately keeping track of each pure contribution, would require $\mathcal{O}(M\varsigma^{-2})$. Surprisingly, such cost can be bypassed whenever the reconstruction is performed agnostically: without keeping track of which pure state yielded each outcome. In such case, i.e. when the mixture is generated by sampling from pure states, $p(1_a\cap1_b)$ can be obtained as the statistical sample average of a random variable, with outcome $\text{Tr}[U_{H}\rho_iU^{\dagger}_{H} \Pi_a \otimes \Pi_b]$ and probability $w_i$ for $i=0,\ldots,M-1$. From the Hoeffding’s inequality~\citep{hoeffding1963probability}, estimating $p(1_a\cap1_b)$ with uncertainty $\varepsilon$ and confidence level $1-\delta$ requires $\mathcal{O}(\varepsilon^{-2}\log(2/\delta))$ photons. Namely, the cost of a single classification instance with a quantum optical shallow network does not scale either with the input size $N$ or the number of hidden neurons $M$. A speedup $\mathcal{O}(C) \to \mathcal{O}(Q)$ is called exponential when $Q \propto \log C$, with $C$ and $Q$ denoting the protocols' resources. Similarly, speedups such that $Q > \log C$ and $Q < \log C$ are called subexponential and superexponential, respectively. In the latter case, no exponential map can pullback the new cost to the original one. Using constant resources, i.e.~$\mathcal{O}(1) < \mathcal{O}(\log(MN)) < \mathcal{O}(MN)$, our protocol asymptotically shows a superexponential speedup over its classical counterpart for a single inference instance. See \cref{app:resources} for a rigorous derivation.

We discuss an alternative encoding strategy for the hidden parameters. We replace the mixture of \cref{eq:MixedState} with a superposition of $M$ states $\ket{W_{\lambda_i}}$, which reads $\rho_{\U_{\lambda}} = \sum_{ij} w_i w_j \ket{W_{\lambda_i}}\!\bra{W_{\lambda_j}}$. In this case, measuring the rate of coincidences gives $f_{w W}(\I) = |\sum_{i}w_i\langle \I , W_{\lambda_i} \rangle|^2$. This coherently provides a single classification instance, still with constant resources. If compared to the mixture, the square absolute value is applied after the scalar product with $w_i$: the model can be cast into a quadratic regression protocol by a simple rotation in the parameters space, without exhibiting universal approximation capabilities. In this case, its performances do not depend on the number of neurons in the hidden layer, making it unfeasible when learning complex patterns. See \cref{app:models,app:training} for a mathematical discussion of this protocol.
\begin{figure*}
	\centering
	\includegraphics[width = .75 \textwidth]{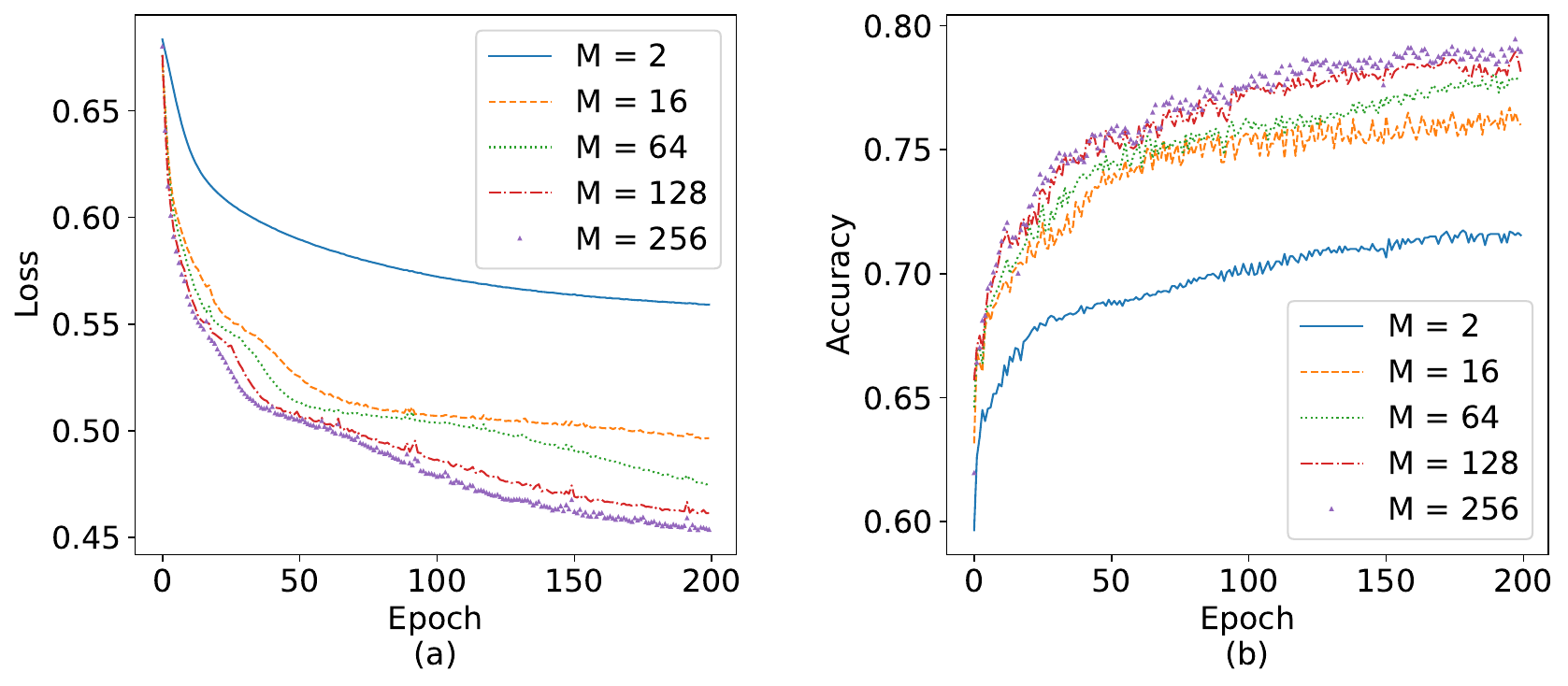}%
	\caption{\label{fig:Training}History plot of the quantum optical shallow network, when classifying images of planes and dogs from the CIFAR-10 validation dataset. The optical network is modelled by an amplitude modulated hidden state of $1024$ pixels. Each line is associated to a different number $M$ of neurons of the hidden layer, i.e.~a combination of multiple hidden parameters into a mixed state (a trainable convex combination of $M$ pure states). Optimization is performed with Adam for $200$ epochs, batch size $1000$ and learning rate $0.03$. (a) Binary cross-entropy versus the number of training epochs. (b) Accuracy versus the number of training epochs. Increasing the number of hidden neurons leads to improved asymptotic accuracy.}
\end{figure*}

Finally, we discuss the possibility of retrieving an advantage by using mixed states to completely dequantize our model. This means replacing the neural network with a perceptron, with parameters randomly changed when sampling over the same input (according to the probability mass function of the mixed state). This protocol mathematically corresponds to our method, but it has no advantage. In our model, sampling introduces only a constant overhead (see the Hoeffding's inequality in \cref{app:resources}) which multiplies the original $\mathcal{O}(1)$ complexity of the Hong-Ou-Mandel neuron: it preserves the superexponential speedup. The quantum-inspired model does not bypass the perceptron's cost of $\mathcal{O}(N)$, which also increases when accounting for Monte-Carlo sampling.

\section{TRAINING AND CLASSIFICATION}
In this section, we discuss the training of the set of parameters $\theta=\{w,W_\lambda,\beta\}$. Then, we present simulations of our model applied to the classification of images.

We consider classification tasks in which approximators, according to \cref{eq:UniversalApproximator}, are trained to draw a boundary between binary classes that are (at least partially) separable in the feature space. Let $D$ be a training set, that is, a collection of data samples $\{\I_s\}$ in the feature space, e.g.~images represented as points whose coordinates correspond to pixel intensities, with known labels $\{y_s \in \{0,1\}\}$. Each training sample is then individually and sequentially fed into the input branch of the interferometer. Classes predicted from the rate of coincidences are compared with the target ones, using the binary-cross entropy $H$ as loss function. Taking into account the normalization of the hidden and the output weights, as well as the positivity of the latter, we describe training as a constrained optimization problem
\begin{gather}	
	\!\min_{\theta} \sum_{s \in D} H(y_s, F_{s\theta}) \\ 
	\begin{split} \text{subject to} \ \ 
	&\!\textstyle \sum_{i} w_i = 1 \ , \ w_i \geq 0 \ \forall i \\ 
	&\|W_{\lambda_i}\|^2 = 1 \ \forall i \ .
	\label{eq:Constraints}
	\end{split}	
\end{gather}
Computing the loss function derivatives, we optimize $H$ via stochastic (mini-batch) gradient descent~\citep{robbins1951stochastic}. Parameters are updated according to the rule
\begin{equation}
    \theta \to \theta - \frac{\eta}{|B|} \sum_{s\in B_l} \nabla_\theta H\left(y_{s},\F_{s\theta}\right) \ ,
    \label{eq:GradientDescent}
\end{equation}
with $\eta$ called learning rate. At each epoch, i.e.~a full pass through the dataset, the update rule is subsequently applied to $\{B_l \subseteq D\}$, called batches, i.e.~elements of a random partition of $D$ into subsets of cardinality $|B|$.

Gradient descent can be constrained by basically two means. On the one hand, one can enforce normalization and positivity directly in \cref{eq:Mixture}, modifying the response function and the gradients accordingly. This approach resembles the paradigm of weight normalization, commonly adopted in classical neural networks~\citep{salimans2016weight}. On the other hand, such conditions can be continuously enforced, by manually constraining the parameters after each application of \cref{eq:GradientDescent}. Although computationally less expensive, this last approach provides a limited exploration of the parameter space, dampening the model expressivity as well as training convergence. See \cref{app:training} for a mathematical discussion that takes into account constraints. We consider a specialization of those results to the case where both the input and the hidden branches are modelled by two amplitude-modulated SLMs, enforcing the constraints after each epoch. Applying the chain rule to \cref{eq:GradientDescent}, the only two model-dependent contributions in $\nabla_\theta H$ are $\nabla_w f$ and $\nabla_{\lambda_i} f$. We obtain $(\nabla_{w}f)_c = \sum_{j=0}^{N-1} |(W_{\lambda_c})_jx_j|^2$ and $(\nabla_{\lambda_{c}}f)_d = 2w_c \sum_{j=0}^{N-1} x_d(W_{\lambda_c})_jx_j$, where $c=0,\ldots,M-1$ and $d=0,\ldots,N-1$. Finally, we reduce the number of constraints in \cref{eq:Constraints}, by keeping track of the output weights normalization $K = \sum_i w_i$, post-processing the output as $\tilde{F}_{\theta} = \sigma(Kf_{w W}(\I) + \beta)$. Despite the increased computational training cost (but not the optical one), this strategy improves the model convergence.

For our simulations, we considered three benchmark datasets: the MNIST dataset~\citep{data:mnist}, consisting of 28 × 28 images of handwritten digits (0-9); the Fashion-MNIST dataset~\citep{data:fashionmnist}, which contains 28x28 grayscale images of clothing items such as T-shirts, pullovers, and bags; and the CIFAR-10 dataset~\citep{data:cifar10}, composed of 32 × 32 color images spanning 10 different categories, including cats, dogs and planes. To ensure consistent input dimensions, we zero-padded the MNIST and Fashion-MNIST images to 32×32 pixels and converted CIFAR-10 images to grayscale (color channels can be encoded by multiple sets of trainable parameters, either executed sequentially on the same apparatus or in parallel on different devices). Each input image is directly encoded in an amplitude-modulated SLM $\hat{\I}(r) = \sum_{j}x_j \delta(r-r_j)$, with $x_j$ the normalized pixel values. We use the binary cross-entropy as loss function and the Adam optimizer, namely a stochastic mini-batch gradient descent accelerated with momentum~\citep{kingma2014adam}. We employ Glorot initialization for both the hidden and output weights~\citep{glorot2010understanding}. Model performance is evaluated in terms of classification accuracy on the test set, defined as the ratio between correctly classified samples and the total number of data samples. The dataset is split into training and test subsets using an 85:15 ratio.
\begin{table}[t]
	\centering
	\def\arraystretch{1.5}
	\setlength\tabcolsep{5pt}
        \begin{tabular}{|l|c|c|c|}
            \cline{2-4}
            \multicolumn{1}{c|}{\multirow{2}{*}{\parbox{1.5cm}{Accuracy\\(Best) [\%]}}} & \multirow{2}{*}{MNIST} & \multirow{2}{*}{\parbox{1.5cm}{\centering Fashion\\MNIST}} & \multirow{2}{*}{CIFAR} \\ \multicolumn{1}{c|}{} &&& \\ \hline
 	      M =  2 & 99.76 & 96.05 & 71.20 \\ \hline
 	      M = 16 & 99.90 & 96.70 & 76.25 \\ \hline
              M = 64 & 99.95 & 96.85 & 77.05 \\ \hline
              M = 128 & 99.95 & 96.90 & 78.30 \\ \hline
              M = 256 & 99.95 & 96.90 & 78.70 \\ \hline
	\end{tabular}	
\caption{\label{tab:QuantumAccuracy}Classification accuracy for the quantum optical shallow network, for different numbers $M$ of hidden neurons, using the datasets MNIST (zeros and ones), Fashion-MNIST (T-shirts/tops and pullovers), and CIFAR-10 (planes and dogs). Model trained with Adam for $150$ epochs, batch size $1000$, and learning rate $0.03$. Increasing the number of neurons improves the asymptotic accuracy.}
\end{table}
\begin{table}[t]
	\centering
	\def\arraystretch{1.5}
	\setlength\tabcolsep{5pt}
	\begin{tabular}{|l|c|c|c|}
            \cline{2-4}
            \multicolumn{1}{c|}{\multirow{2}{*}{\parbox{1.5cm}{Accuracy\\(Best) [\%]}}} & \multirow{2}{*}{MNIST} & \multirow{2}{*}{\parbox{1.5cm}{\centering Fashion\\MNIST}} & \multirow{2}{*}{CIFAR} \\ \multicolumn{1}{c|}{} &&& \\ \hline
 	    M = 2  & 99.95 & 96.90 & 74.70 \\ \hline
 	    M = 16 & 99.95 & 96.80 & 78.45 \\ \hline
            M = 64 & 99.91 & 96.75 & 78.35  \\ \hline
            M = 128 & 99.95 & 96.80 & 77.65  \\ \hline
            M = 256 & 99.91 & 96.70 & 77.60 \\ \hline
	\end{tabular}	
\caption{\label{tab:ClassicalAccuracy}Classification accuracy for the classical (unconstrained) shallow neural network, for different numbers $M$ of hidden neurons, using the same datasets of~\cref{tab:QuantumAccuracy}. Trained with mini-batch gradient descent (without Adam) for $1000$ epochs, batch size $1000$, and learning rate $0.03$.}
\end{table}

In \cref{fig:Training}, we show the history plot when distinguishing between planes and dogs in the CIFAR-10 dataset. With $M=2$ hidden neurons, our model reaches an asymptotic accuracy of $71\%$ after $200$ epochs. Increasing the number of hidden neurons to $M=256$ results in a saturated accuracy of approximately $78\%$. This behavior is consistent with \cref{eq:UniversalApproximator}, which states that increasing the complexity of the hidden layer enhances the network's ability to approximate the decision boundary. In \cref{tab:QuantumAccuracy}, we report the best (i.e. maximal) test accuracy achieved across all datasets as a function of the number of hidden neurons ($M=2, 16, 64, 128, 256$). In~\cref{tab:ClassicalAccuracy}, we present the same analysis for unconstrained classical neural networks with comparable architecture and complexity. Expectedly, the classical and quantum models show similar accuracy under the same parameter count. A thorough investigation is reported in \cref{app:classical}, where we discuss how accuracy is affected when the constraints of the quantum model ($L^2$ and $L^1$ normalizations) are enforced on the classical one.

\section{CONCLUSIONS}
In this work, we designed a quantum optical analogue of a shallow neural network. We mathematically proved the analogy between our protocol and its classical counterpart, demonstrating its universal approximation capabilities (provided that the number of hidden neurons is sufficient to counteract the normalization constraints). Increasing the number of neurons leads to improved expressivity, without modifying the number of resources: a superexponential speedup in inference, with comparable accuracy over classical shallow neural networks. This opens new possibilities for inference tasks in low photon-number conditions, or for evaluating kernel methods in embedded devices with limited computational power.

Despite supporting an arbitrary number of hidden neurons, our architecture is still constrained to scalar outputs. Increasing the output dimension would include new possible applications, from multinomial classification tasks to multivariate regressions. For example, this can be achieved by employing alternative encoding methods, multi-port optical components in place of the beam splitter, or partial photon-number-resolving detectors to encode multiple output nodes in multiple regions of limited spatial sensitivity. Another possible future direction is a rigorous mathematical characterization of the universal approximation capabilities of the proposed model, taking into account the underlying physical constraints and normalization conditions.

Beyond these extensions, the proposed approach is particularly well suited to inference tasks such as binary hypothesis testing in imaging, where the objective is to detect the presence of a target or an anomaly, rather than reconstruct a full image. Such scenarios arise in biological sensing and remote detection under stringent photon constraints, due to eye-safety requirements, weak signals, or limited acquisition times, where conventional pixel-resolved imaging is inherently inefficient.

\section*{ACKNOWLEDGEMENTS}
S.R. acknowledges support from the PRIN MUR Project 2022SW3RPY. A.R.M. acknowledges support from the PNRR MUR Project PE0000023-NQSTI. C.M. acknowledges support from the National Research Centre for HPC, Big Data and Quantum Computing, PNRR MUR Project CN0000013-ICSC. L.M. acknowledges support from the PRIN MUR Project 2022RATBS4. SL: This material is based upon work supported by, or in part by, the U. S. Army Research Laboratory and the U. S. Army Research Office under contract/grant number W911NF2310255, and by DoE under contract, DE-SC0012704.

\section*{CODE AVAILABILITY}
The underlying code that generated the data for this study is openly available in GitHub \citep{rep:QON}.

\appendix
\renewcommand{\thefigure}{S\arabic{figure}}
\setcounter{figure}{0}

\renewcommand{\thetable}{S\arabic{table}}
\setcounter{table}{0}

% \clearpage

\begin{widetext}
\section{HONG-OU-MANDEL COINCIDENCES\label{app:models}}
In this section, we compute the rate of coincidences at the output of a Hong-Ou-Mandel interferometer, when feeding a generic (single-photon) density operator $\rho$ into one branch, and a pure state $\ket{\I}$ in the other. Adopting the monochromatic approximation, we assume that the wavefront propagates along definite-sign $z$-directions only, i.e. by considering its transversal degree of freedom only. In the input branch, we consider the single-photon state 
\begin{equation}
	\ket{\I} = \int \dif^2k \ \I_\omega(k) a^\dagger_\omega(k) \ket{0} \ ,
\end{equation}
where $\hat{\I}_\omega(k) = (\hat{\Phi}_\omega*\hat{\O}) \hat{\mathcal{L}}_{d}$ is the single-photon spectrum (with source $\Phi$) that encodes the object shape $\O$, after a linear optical system with transfer function $\mathcal{L}_{d}$, with $*$ the convolution operator. In the hidden branch, we consider the single-photon density operator
\begin{equation}
    \rho_{\U_\lambda} = \int \dif^2k_1 \dif^2k_2 \ \U_\lambda(k_1,k_2)b_\omega^{\dagger}(k_1)\ket{0}\!\bra{0}b_\omega(k_2) \ ,
\end{equation}
with $\lambda$ a collection of trainable parameters. We write the input-hidden bipartite density operator as
\begin{equation}
	\ket{\I}\!\bra{\I} \otimes \rho_{\U_\lambda} = \int \dif^2k_1 \dif^2k_2 \dif^2k_3 \dif^2k_4 \ \I(k_1) \I^*(k_2) \U_\lambda(k_3,k_4) a^\dagger(k_1) b^\dagger(k_3) \ket{0}\!\bra{0} a(k_2) b(k_4) \ ,
	\label{eq:RhoBipartite}
\end{equation}
From now on, we omit the frequency subscript $\omega$. The balanced beam splitter acts on each mode as
\begin{equation}
	\begin{cases}
		a^\dagger \to \frac{1}{\sqrt{2}}\left( a^\dagger + b^\dagger \right) \\
		b^\dagger \to \frac{1}{\sqrt{2}}\left( a^\dagger - b^\dagger \right)
	\end{cases} \ ,
	\label{eq:BeamSplitterOperation}
\end{equation} 
with output density operator
\begin{equation}
	 \rho = \frac{1}{4} \int \prod_{i=1}^4 \dif^2k_i \ \I(k_1) \I^*(k_2) \U_\lambda(k_3,k_4) \big[ a^\dagger(k_1) + b^\dagger(k_1) \big]\big[ a^\dagger(k_3) - b^\dagger(k_3) \big] \ket{0}\!\bra{0} \big[ a(k_2) + b(k_2) \big]\big[ a(k_4) - b(k_4) \big] \ .
	\label{eq:RhoBipartiteOutput}
\end{equation}
Detection of mode $m \in \{a,b\}$ is described by the projector $\Pi_{m} = \int \dif^2k \ m^\dagger(k)\ket{0}\!\bra{0}m(k)$. The rate of coincidences, i.e. the probability that one and only one photon is detected in each mode, reads
\begin{gather}
	 p(1_a \cap 1_b) = \Tr[\rho\Pi_a \otimes \Pi_b] \ , \\
	\text{with} \ \Pi_a \otimes \Pi_b = \int \dif^2k_5 \dif^2k_6 \ a^\dagger(k_5) b^\dagger(k_6) \ket{0}\!\bra{0} a(k_5) b(k_6) \ .
\end{gather}
By substitution of \cref{eq:RhoBipartiteOutput}, we get
\begin{equation}
	p(1_a \cap 1_b) = \frac{1}{4}\int \prod_{i=1}^6 \dif^2k_i \ \I(k_1) \I^*(k_2) \U_\lambda(k_3,k_4) A(k)\ , 
	\label{eq:CoincStep}
\end{equation}
where $k = (k_1,k_2,\ldots,k_6)$ and
\begin{equation}
\begin{split}
		A(k) &= \bra{0} a(k_5)b(k_6)\big[ a^\dagger(k_1) + b^\dagger(k_1) \big]\big[ a^\dagger(k_3) - b^\dagger(k_3) \big] \ket{0} \! \bra{0} \big[ a(k_2) + b(k_2) \big]\big[ a(k_4) - b(k_4) \big] a^\dagger(k_5)b^\dagger(k_6) \ket{0} \\
			&= \left[\delta(k_1 - k_6)\delta(k_3 - k_5) - \delta(k_1 - k_5)\delta(k_3 - k_6)\right] \left[\delta(k_2 - k_6)\delta(k_4 - k_5) - \delta(k_2 - k_5)\delta(k_4 - k_6)\right] \ .
\end{split}
\end{equation}
After integrating on $k_5$ and $k_6$, we obtain
\begin{gather}
	p(1_a \cap 1_b) = \frac{1}{2}\int \prod_{i=1}^4 \dif^2k_i \ \I(k_1) \I^*(k_2) \U_\lambda(k_3,k_4) \left[ \delta(k_1-k_2)\delta(k_3-k_4) - \delta(k_1-k_4)\delta(k_2-k_3)\right] \\
	 \Rightarrow \ p(1_a \cap 1_b) = \frac{1}{2}\left[\| \I \|^2 \Tr[\rho_{\U_\lambda}] -  \int \dif^2k_1 \dif^2k_2 \ \I(k_1) \I^*(k_2) \U_\lambda(k_2,k_1) \right] \ ,
	 \label{eq:GeneralCoincidence}
\end{gather}
with $\| \cdot \|$  denoting the $L^2$-norm. We define $C = \| \I \|^2 \Tr[\rho_{\U_\lambda}]$, an additional hyperparameter, taking into account optical losses that can prevent normalization. We discuss \cref{eq:GeneralCoincidence} in several scenarios, i.e. when $\rho_{\U_\lambda}$ describes a single pure state, a superposition or a mixed state.

\subparagraph{Pure state}
In this section, we derive the response function of a single artificial neuron. Let $\rho_{\U_\lambda}$ be a (single) pure state, namely $\rho_{\U_\lambda} = \ket{W_\lambda}\!\bra{W_\lambda}$ with $\U_\lambda(k_1,k_2) = W_\lambda(k_1)W_\lambda^*(k_2)$. By substitution in \cref{eq:GeneralCoincidence}, we get  
\begin{equation}
	p(1_a \cap 1_b) = \frac{1}{2}\left[\| \I \|^2 \| W_\lambda \|^2 -  |\langle \I , W_\lambda \rangle|^2 \right] \ , 
\end{equation}
where $\langle \cdot , \cdot \rangle$ denotes the inner product in $L^2$. With $\Tr[\rho_{\U_\lambda}] = \| W_\lambda \|^2$ and $f_{W}(\I) = C - 2p(1_a \cap 1_b)$, this formula reads
\begin{equation}
	f_{W}(\I) = |\langle \I , W_\lambda \rangle|^2 \ .
        \label{eq:SingleNeuron}
\end{equation} 
This is the response function of a single neuron (with square modulus activation function), already discussed in~\citep{roncallo2025quantum}. In this case, the neuron is provided with no bias and a square absolute value activation function, with (trainable) parameters encoded in the pure state $\ket{W_\lambda}$. \cref{eq:SingleNeuron} requires $\mathcal{O}(1)$ computational resources and $\mathcal{O}(\varsigma^{-2})$ photons, for a single classification instance with uncertainty $\varsigma$. This is a superexponential speedup over a classical artificial neuron, whose cost scales linearly in the number of input features (both optically and computationally). See~\citep{roncallo2025quantum} for a thorough analysis of the classical scenario, and \cref{app:resources} for a discussion about optical resources. 

\subparagraph{Mixture}
In this section, we describe the response function of a neural network with $M$ hidden neurons. Consider a mixed state $\rho_{\U_\lambda}$, namely a convex combination of pure states $\{\ket{W_{\lambda_{i}}} : i = 0,\ldots,M-1\}$ with probabilities $\{w_i \geq 0 : i = 0,\ldots,M-1\}$. Hence
\begin{gather}
    \U(k_1,k_2) = \sum_{i=0}^{M-1} w_i W_{\lambda_{i}}(k_1)W_{\lambda_{i}}^*(k_2) \ , \\
    \text{where} \sum_{i=0}^{M-1} w_i \|W_{\lambda_{i}}\|^2 = 1 \ . \label{eq:MixtureNormalization}
\end{gather}
By substitution in \cref{eq:GeneralCoincidence}, we get  
\begin{equation}
    p(1_a \cap 1_b) = \frac{1}{2}\left[\| \I \|^2 \Tr[\rho_{\U_\lambda}] -  \sum_{i=0}^{M-1}w_i|\langle \I , W_{\lambda_{i}} \rangle|^2 \right] \ ,
\end{equation}
where $\Tr[\rho_{\U_\lambda}] = \sum_{i}w_i\|W_{\lambda_{i}}\|^2$. Assuming the normalization of $\ket{W_{\lambda_{i}}}$, i.e. $\|W_{\lambda_{i}}\|^2 = 1 \ \forall i$, \cref{eq:MixtureNormalization} simplifies to $\sum_i w_i = 1$. Writing the last equation with respect to $f_{wW}(\I) = C - 2p(1_a \cap 1_b)$, we obtain
\begin{equation}
    f_{wW}(\I) = \sum_{i=0}^{M-1}w_i|\langle \I , W_{\lambda_{i}} \rangle|^2 \ .
    \label{appeq:Mixture}
\end{equation}
This is precisely the output of a neural network with a single hidden layer of $M$ neurons, i.e.~a universal approximator according to Hornik-Stinchcombe-White theorem~\citep{hornik1989multilayer}, with no biases and a square modulus activation function, constrained by the normalization of \cref{eq:MixtureNormalization} and the positivity of the output weights $w_i \geq 0$.

\subparagraph{Superposition}
Consider $\rho_{\U_\lambda}$ as a superposition of pure states $\{\ket{W_{\lambda_{i}}} : i = 0,\ldots,M-1\}$ with amplitudes $\{w_i \in \mathbb{C} : i = 0,\ldots,M-1\}$. Namely
\begin{gather}
    \U(k_1,k_2) = \sum_{i,j=0}^{M-1} w_i w^*_j W_{\lambda_{i}}(k_1)W_{\lambda_{j}}^*(k_2) \ , \\
    \text{where} \ \Tr[\rho_{\U_\lambda}] = \sum_{i,j=0}^{M-1}w_iw^*_j\int \dif^2 k \  W_{\lambda_{i}}(k)W_{\lambda_{j}}^*(k) = 1 \ . \label{eq:SuperpositionNormalization}
\end{gather}
By substitution in \cref{eq:GeneralCoincidence}, we get  
\begin{gather}
	p(1_a \cap 1_b) = \frac{1}{2}\Big[\| \I \|^2 \Tr[\rho_{\U_\lambda}] -  \sum_{i,j=0}^{M-1}w_iw^*_j\langle \I , W_{\lambda_{i}} \rangle\langle W_{\lambda_{j}} , \I \rangle \Big] \  , \\
	f_{w W}(\I) = \Big|\sum_{i=0}^{M-1}w_i\langle \I , W_{\lambda_{i}} \rangle\Big|^2 \ , \label{eq:Superposition}
\end{gather}
with $f_{wW}(\I) = C - 2p(1_a \cap 1_b)$. From now on, we assume that $w_i \in \mathbb{R}$. This model represents an alternative encoding scheme for the hidden parameters, namely a coherent implementation that has the same resource costs of the mixture. Several issues arise in this scenario. Mathematically, \cref{eq:Superposition} is not a universal approximator, in the sense of Hornik-Stinchcombe-White theorem~\citep{hornik1989multilayer}, since it describes $M$ hidden neurons but with linear activation functions. A single instance of \cref{eq:Superposition} can be reparametrized, i.e. collapsed, to that of a single neuron. In quantum mechanical terms: superposition can be viewed as a change of basis on the experimental apparatus implemented by a single pure component $\ket{W_{\lambda_i}}$. This implies limited learning capabilities, independently on the network complexity $M$. Such limitations can also be confirmed numerically. In our simulations, this model did not exhibit the same accuracy scalability of the mixture (with respect to the number $M$ of hidden neurons).

\section{LOSS FUNCTION DERIVATIVES\label{app:training}}
In this section, we discuss training for the quantum optical shallow network. In particular, we compute the derivatives associated to the implementations discussed in the previous section. First, we describe gradients in general terms. Then, we specialize to \cref{appeq:Mixture,eq:Superposition}. Let us define the set of training parameters $\theta = \{w, W_\lambda, \beta\}$, where $w \in \mathbb{R}^M$ are the output weights, $\beta \in \mathbb{R}$ the output bias and $\{W_{\lambda_i} \in \mathbb{R}^N : i=0,\ldots,M-1\}$ the weights of the hidden layer, i.e. a set of $M$ neurons each with $N$ weights such that $W_\lambda \in \mathbb{R}^{M\times N}$. Training is performed by applying an Adam-optimized mini-batch gradient descent on the binary cross-entropy
\begin{equation}
    H(y_s, F_{s\theta}) = -y_s \log{F_{s\theta}} - (1- y_s) \log{(1-F_{s\theta}}),
\end{equation}
where $y_s \in \{0,1\}$ is the ideal output (i.e. correct class label) of the $s$th data sample of the training dataset $D$, while $F_{s\theta} \in (0,1)$ is the corresponding predicted label, computed as 
\begin{equation}
	F_{s\theta} = g(f_{w W}(\I_s) + \beta) \ ,
\end{equation}
with $g$ the activation function, e.g. the sigmoid $\sigma(x) = 1/[1+\exp(-x)]$, and $\I_s$ the $s$th sample of the dataset. Here, $f_{w W}$ represents the output of the models, given by \cref{appeq:Mixture} or \cref{eq:Superposition}, and obtained from the probabilities of the Hong-Ou-Mandel coincidences. Independently of the model, \cref{eq:GeneralCoincidence} reads 
\begin{equation}
	f_{wW}(\I) = \int \dif^2k_1 \dif^2k_2 \ \I(k_1) \I^*(k_2) \U_\lambda(k_2,k_1) \ .
\end{equation}
We compute the derivative of the loss function by applying the chain rule $\nabla_\theta H = \partial_F H \partial_\xi g \nabla_\theta \xi,$ where $\xi = f_{w W} + \beta$. For each set of parameters we get $\partial_\beta \xi = 1$ and
\begin{equation}
\begin{gathered}
    \nabla_{w} \xi = \nabla_{w} f = \int \dif^2 k_1 \dif^2 k_2 \ \I(k_1) \I^*(k_2) \, \nabla_{w} \U(k_2,k_1) \ , \\
    \nabla_{\lambda_c} \xi = \nabla_{\lambda_c} f = \int \dif^2 k_1 \dif^2 k_2 \ \I(k_1) \I^*(k_2) \, \nabla_{\lambda_c} \U(k_2,k_1) \ ,
    \label{eq:GeneralDerivatives}
\end{gathered}
\end{equation}
where $\nabla_{\lambda_c} \in \{\nabla_{\lambda_0},\ldots, \nabla_{\lambda_{M-1}}\}$ labels the derivative with respect to the $c$th hidden neuron, each with $N$ components labelled by $d$, i.e.~$(\nabla_{\lambda_c}f)_d \in \{(\nabla_{\lambda_c}f)_0,\ldots,(\nabla_{\lambda_c}f)_{N-1}
\}$. Similarly, $(\nabla_{w}f)_c \in \{(\nabla_{w}f)_0,\ldots,(\nabla_{w}f)_{M-1}
\}$. From now on, we guarantee differentiability by assuming that all the parameters are real. Otherwise, the binary cross-entropy (which is real-valued) would not be holomorphic. After computing the loss function derivative, the parameters are updated according to the rule
\begin{equation}
    \theta \to \theta - \frac{\eta}{|B|} \sum_{s\in B_l} \nabla_\theta H\left(y_{s},\F_{s\theta}\right) \ ,
\end{equation}
with $\eta$ called learning rate. At each epoch, the update rule is subsequently applied for each batch $B_l \subseteq D$, obtained by randomly dividing the dataset in partitions of $|B|$ data samples, with cardinality $|B|$. 

\subparagraph{Mixture}
In this section, we explicitly discuss the mixture case. After each epoch, the parameters are updated and normalization needs to be guaranteed by \cref{eq:MixtureNormalization}. We do this, by enforcing such condition directly in $f_{wW}$, modifying the derivative while keeping track of the additional terms. Classically, this resembles the concept of weight normalization~\citep{salimans2016weight}. Hence
\begin{equation}
    \U(k_1,k_2) = \sum_{i=0}^{M-1} \frac{P(w_i)}{\sum_j P(w_j)} \frac{W_{\lambda_{i}}(k_1)W_{\lambda_{i}}^*(k_2)}{\|W_{\lambda_{i}} \|^2} \ ,
\end{equation}
where $P$ is a function enforcing the positivity of the output weights, i.e.~$P(w_i) \geq 0 \,\, \forall i$. For example $P \in \{|\cdot|, \ \sigma, \ \text{ReLU} \}$. Taking the derivative with respect to the $c$th output weight, we get
\begin{align}
    (\nabla_{w}\,\U)_c &= - \sum_{p=0}^{M-1} \frac{\partial_{w_c} P(w_p)}{\sum_j P(w_j)} \sum_{i=0}^{M-1} \frac{P(w_i)}{\sum_j P(w_j)}\frac{W_{\lambda_{i}}(k_1)W_{\lambda_{i}}^*(k_2)}{\|W_{\lambda_{i}} \|^2} + \sum_{i=0}^{M-1} \frac{\partial_{w_c}P(w_i)}{\sum_j P(w_j)}\frac{W_{\lambda_{i}}(k_1)W_{\lambda_{i}}^*(k_2)}{\|W_{\lambda_{i}}\|^2} \\
 &= \frac{P'(w_c)}{\sum_j P(w_j)} \left[ -\U(k_1,k_2) + \frac{W_{\lambda_{c}}(k_1)W_{\lambda_{c}}^*(k_2) }{\| W_{\lambda_{c}} \|^2} \right] \ .
 \label{eq:derwaU}
\end{align}
In the second step, we used that $\partial_{w_c}P(w_i) = \delta_{c i}P'(w_i)$, with $\delta_{c i}$ denoting the Kronecker delta and $P'(w_i) = \partial_{w_i}P(w_i)$. Hence, $\sum_p\partial_{w_c}P(w_p) = P'(w_c)$. By substituting in \cref{eq:GeneralDerivatives}, we obtain
\begin{equation}
     (\nabla_{w}f)_c = \frac{P'(w_c)}{\sum_jP(w_j)}\left[ \ -f_{wW}(\I) + \left|\Big\langle \I , \frac{W_{\lambda_{c}}}{\| W_{\lambda_{c}}\|}\Big\rangle\right|^2 \ \right] \ .
    \label{eq:dwaf}
\end{equation}
Computing the derivative with respect to the hidden parameters, we get
\begin{equation}
	\begin{gathered}
  	(\nabla_{\lambda_c}\U)_d =  \frac{P(w_c)}{\sum_jP(w_j)\|W_{\lambda_{c}} \|^2} \Bigg[-\frac{2W_{\lambda_{c}}(k_1)W_{\lambda_{c}}^*(k_2)}{\|W_{\lambda_{c}} \|^2} \Re[\langle W_{\lambda_{c}},(\nabla_{\lambda_c}W_{\lambda_{c}})_d\rangle] 
  	+ \left[\nabla_{\lambda_c}\left(W_{\lambda_{c}}(k_1)W_{\lambda_{c}}^*(k_2)\right)\right]_d \Bigg] \ .
    \label{eq:derlambdaU}
    \end{gathered}
\end{equation}
By substitution in \cref{eq:GeneralDerivatives}, we finally obtain
\begin{equation}
    (\nabla_{\lambda_c}f)_d = \frac{2P(w_c)}{\sum_jP(w_j)}\text{Re}\left[\Big\langle \I , \frac{(\nabla_{\lambda_c}W_{\lambda_{c}})_d}{\|W_{\lambda_{c}} \|}\Big\rangle \Big\langle\frac{W_{\lambda_{c}}}{\| W_{\lambda_{c}}\|} , \I\Big\rangle - \left|\Big\langle \I , \frac{W_{\lambda_{c}}}{\|W_{\lambda_{c}}\|}\Big\rangle \right|^2 \Big\langle\frac{W_{\lambda_{c}}}{\|W_{\lambda_{c}}\|} , \frac{(\nabla_{\lambda_c}W_{\lambda_{c}})_d}{\|W_{\lambda_{c}} \|}\Big\rangle \right] \ .
    \label{eq:dwaf2}
\end{equation}
As an explicit example, we consider the (amplitude- or phase-modulated) encoding provided by a spatial light modulator (SLM) of $N$ pixels. The mixture can be generated by considering $M$ independent patterns on the SLM, and randomly sampled with probabilities $\{w_i\}$. We denote the set of amplitudes and phases as $\{\alpha_{ij}\}$ and $\{\varphi_{ij}\}$, respectively, where $i = 0, \ldots, M-1$ labels the pure components of the mixture and $j = 0, \ldots, N-1$ each pixel position. Namely
\begin{equation}
    (\lambda_{c})_d = (\alpha_{c d},\varphi_{c d}) \to \hat{W}_{\lambda_{c}}(r)= \sum_{d=0}^{N-1} \alpha_{c d}\exp(i\varphi_{c d})\delta(r - r_d) \ ,
\end{equation}
where $r\in\mathbb{R}^2$ labels the transverse coordinate on the SLM plane. We consider the same representation for the input branch, in which a digital image $\{\I_d\}$ is encoded as $\hat{\I}(r) = \sum_{d}\I_d \delta(r-r_d)$. Hence, the terms in \cref{eq:dwaf,eq:dwaf2} simplify as
\begin{gather}
    \langle W_{\lambda_{c}},(\nabla_{\alpha_c}W_{\lambda_{c}})_d\rangle = \alpha_{c d} \ , \quad \langle W_{\lambda_{c}} , (\nabla_{\varphi_c}W_{\lambda_{c}})_d\rangle = i \alpha_{c d}^2 \ , \label{eq:SLM1} \\
	\langle \I,(\nabla_{\alpha_c}W_{\lambda_{c}})_d \rangle = \I^*_d \exp(i\varphi_{c d}) \ , \quad \langle \I , (\nabla_{\varphi_c}W_{\lambda_{c}})_d \rangle = i \I^*_d \alpha_{c d} \exp(i\varphi_{c d}) \ . \label{eq:SLM2}
\end{gather}
We can set $\varphi_{ij} = 0$ for amplitude encoding, or $\alpha_{ij} = 1/N$ for phase encoding. For example, in the former case we obtain
\begin{equation}
    (\nabla_{\lambda_c}f)_d = \frac{2P(w_c)}{\sum_jP(w_j)\|W_{\lambda_{c}} \|^2}\text{Re}\left[ \I^*_d \langle W_{\lambda_{c}} , \I \rangle - \alpha_{c d} \left|\Big\langle \I , \frac{W_{\lambda_{c}}}{\|W_{\lambda_{c}}\|}\Big\rangle \right|^2 \right] \ .
\end{equation}

\subparagraph{Superposition}
We perform a similar analysis on \cref{eq:Superposition}. We enforce the normalization condition directly in $f_{wW}$, modifying the derivative while keeping track of the additional terms. Moreover, we assume that the output weights are real, i.e. $w_i \in \mathbb{R}$, guaranteeing the differentiability of the loss function. Hence, the hidden state reads 
\begin{equation}
	\U(k_1, k_2) = \frac{\sum_{ij=0}^{M-1}w_iw_jW_{\lambda_{i}}(k_1)W_{\lambda_{j}}^*(k_2)}{\sum_{pq=0}^{M-1} w_pw_q\int \dif^2 k \ W_{\lambda_{p}}(k)W_{\lambda_{q}}^*(k)} \ ,
\end{equation}
where we employed the normalization condition of \cref{eq:SuperpositionNormalization}. As for the mixture, we compute the derivatives of $\U$ and $f$ with respect to the output weights $w$, yielding
\begin{gather}
    (\nabla_{w}\,\U)_c = \sum_{i=0}^{M-1} \frac{w_i}{\Tr[\rho_{\U_\lambda}]}  \left[-2 \int \dif^2 k \, \Re[W_{\lambda_{c}}(k)W_{\lambda_{i}}^*(k)] \U(k_1,k_2) +  W_{\lambda_{c}}(k_1) W_{\lambda_{i}}^*(k_2) + W_{\lambda_{i}}(k_1) W_{\lambda_{c}}^*(k_2) \right] \ , \nonumber \\
    \Rightarrow (\nabla_{w}f)_c = \sum_{i=0}^{M-1} \frac{2w_i}{\Tr[\rho_{\U_\lambda}]} \Re[-\langle W_{\lambda_{i}},W_{\lambda_{c}} \rangle f_{wW} + \langle \I , W_{\lambda_{c}}\rangle \langle W_{\lambda_{i
    }} ,\I\rangle] \ .
\end{gather}
Similarly, the derivative with respect to the set of hidden weights reads 
\begin{gather}
    (\nabla_{\lambda_c}\U)_d = \sum_{i=0}^{M-1} \frac{w_i}{\Tr[\rho_{\U_\lambda}]}\left[-2w_c \int \dif^2 k \, \Re[(\nabla_{\lambda_c}W_{\lambda_{c}})_d(k)W_{\lambda_{i}}^*(k)]\U(k_1,k_2)+ \textstyle\sum_{j} w_j \left[\nabla_{\lambda_{c}}\left(W_{\lambda_{i}}(k_1)W_{\lambda_{j}}^*(k_2)\right)\right]_d\right] \ ,  \nonumber \\
    \Rightarrow (\nabla_{\lambda_c}f)_d = \sum_{i=0}^{M-1} \frac{2w_iw_c}{\Tr[\rho_{\U_\lambda}]} \Re[-\langle W_{\lambda_{i}}, (\nabla_{\lambda_c}W_{\lambda_{c}})_d\rangle f_{wW} + \langle \I , (\nabla_{\lambda_c}W_{\lambda_{c}})_d\rangle \langle W_{\lambda_{i}},\I\rangle] \ .
\end{gather}
We can simplify this equation, by employing the SLM model discussed in the previous section. For amplitude encoding, i.e. $\varphi_{ij}=0$, \cref{eq:SLM1,eq:SLM2} imply
\begin{equation}
	(\nabla_{\lambda_c}f)_d = \sum_{i=0}^{M-1} \frac{2w_iw_c}{\Tr[\rho_{\U_\lambda}]} \Re[\I^*_d\langle W_{\lambda_{i}},\I\rangle -\alpha_{cd} f_{wW}] \ .
\end{equation}

\section{STATISTICAL SAMPLE COMPLEXITY AND RESOURCES\label{app:resources}}
In principle, our protocol requires only two photons for a single classification instance, namely $\mathcal{O}(1)$ constant resources. However, predicted classes are obtained from the rate of coincidences, i.e.~a probability to be estimated by gathering multiple statistical samples, which correspond to different experimental repetitions. In this section, we discuss the number of statistical samples, i.e.~photons, needed to estimate $f_{wW}(\I)$, i.e.~the output of the quantum optical shallow network with a mixed hidden state. We first review the cost of estimating the rate of coincidences for a single pure hidden state $\ket{W_{\lambda_i}}$. Then, we generalize our results when taking into account the sampling cost for a mixture of $M$ pure states. 
\begin{figure}[t]
	\centering
	\includegraphics[width = .75 \textwidth]{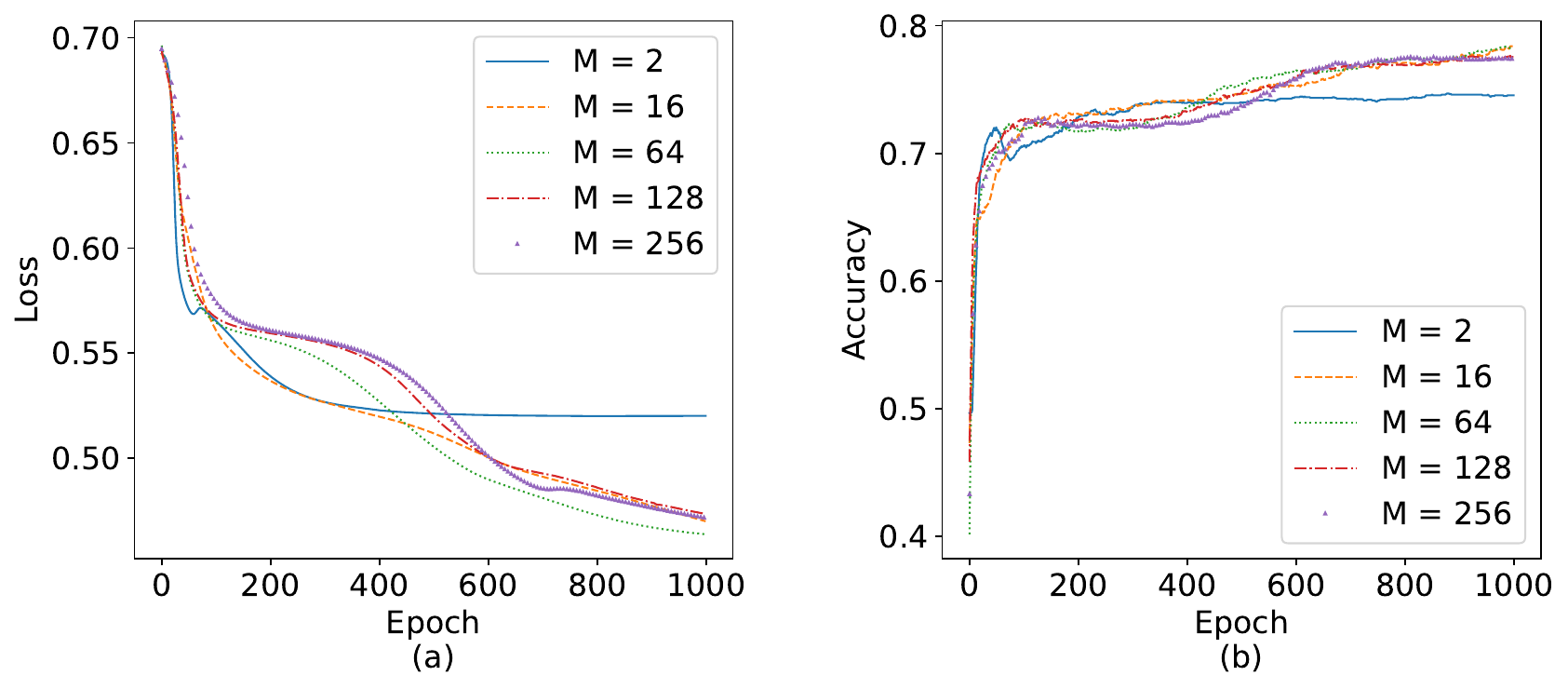}%
	\caption{\label{fig:TrainingClassical}History plot of an unconstrained classical shallow network, when classifying images of planes and dogs from the CIFAR-10 validation dataset. Each line is associated to a different number $M$ of hidden neurons. Optimization is performed for $1000$ epochs, batch size $1000$ and learning rate $0.03$. (a) Binary cross-entropy versus the number of training epochs. (b) Accuracy versus the number of training epochs.}
\end{figure}

In our estimations, we leverage the concept of sample complexity~\citep{arunachalam2018optimal,cheng2025invitation}. In practice, we estimate the number of statistical samples $n$ needed to predict the target label with a given uncertainty. Consider the $i$th pure component $\rho_i$ of the mixed state before the measurement. For an input state with $N$ features (e.g. a digital image of $N$ pixels), the coincidence measurement corresponds to the outcome of a single optical neuron with $N$ parameters, i.e. $|\langle \I , W_{\lambda_i} \rangle|^2 = 1 - 2p_i$, with $p_i \in [0,1/2]$. Let $n_i$ be the number of photons needed to estimate $p_i$. The coincidence probability follows a binomial distribution (with success-failure outcomes corresponding to the presence-absence of a coincidence event). Using the empirical frequencies $f_i$, the probability can be estimated as 
\begin{equation}
	p_i = f_i \pm \varepsilon_i \
	\text{with} \ \varepsilon_i = 2\sqrt{\frac{f_i(1-f_i)}{n_{i}}} \ ,
	\label{appeq:Binomial}
\end{equation}
where the factor $2$ approximately guarantees a $0.95$ confidence level. Since $f_i(1-f_i)\leq 1/4$, inverting this relation yields $n_i \geq \varepsilon_i^{-2}$. Requiring $\varsigma = \varepsilon_i \ \forall i$, the sample complexity of each pure state is $\mathcal{O}(\varsigma^{-2})$. Such bound does not depend on the number of features (e.g. pixels) encoded in $\rho_i$. 

We now derive the total cost of estimating the rate of coincidences $p(1_a \cap 1_b)$ of the mixed state $\rho = \sum_i w_i \rho_i$, with probabilities $w_i \geq 0$. Intuitively, a reconstruction of $p(1_a \cap 1_b)$ that separately keeps track of each pure contribution would cost $\mathcal{O}(M\varsigma^{-2})$, namely $M$ independent experimental repetitions (each time with a different $\rho_i$). This is the same cost of reconstructing a multinomial probability distribution, which scales linearly with the number of possible outcomes (bins), i.e. $\mathcal{O}(M)$. Surprisingly, this is not the case since we are interested only in estimating the statistical sample average of the rate coincidences sampled on $\rho$ (without tracking which pure state yielded each outcome). Indeed, $f_{wW}(\I) = 1-2p(1_a \cap 1_b)$, with 
\begin{gather}
	p(1_a \cap 1_b) = \sum_{i=0}^{M-1}w_i p_i \
	\text{with} \ \textstyle\sum_{i} w_i = 1 \ \text{and} \ w_i \geq 0 \ \forall i \ .
\end{gather}
Namely, the output of the optical shallow network can be estimated from the statistical sample average of a random variable, bounded in $[0,1/2]$, with outcomes $p_0,\ldots,p_{M-1}$ and probabilities $w_0,\ldots,w_{M-1}$. We can estimate the sample complexity using the Hoeffding's inequality (also related to the Chernoff bound)~\citep{hoeffding1963probability} .
\begin{theorem}[Hoeffding’s inequality]\label{th:Theorem}
Let $X_0,\ldots,X_{n-1}$ be $n$ independent random variables (with finite first and second moments) bounded as $X_i \in [0,1] \ \forall i$. For the statistical sample average $\bar{X} = \sum_i X_i / n$ with expectation value $\mathbb{E}[\bar{X}]$ and for uncertainty $\varepsilon$ with $0<\varepsilon < 1 - \mathbb{E}[\bar{X}]$, we get 
\begin{equation}
	\Pr(\bar{X} - \mathbb{E}[\bar{X}] \geq \varepsilon) = \Pr(\textstyle\sum_i X_i - n\mathbb{E}[\bar{X}] \geq n \varepsilon)\leq \exp(-2n\varepsilon^2) \ .
\end{equation}
\end{theorem}
In our case, the random variables are identically distributed, yielding $\mathbb{E}[\bar{X}] = \mathbb{E}[X] = p(1_a \cap 1_b)$. For the confidence level $1-\delta$, with $\delta = \Pr(|\bar{X} - p(1_a \cap 1_b)| \geq \varepsilon)$, we get 
\begin{align}
    \delta &= \Pr(\bar{X} - p(1_a \cap 1_b) \geq \varepsilon) + \Pr(-\bar{X} + p(1_a \cap 1_b) \geq  \varepsilon) \\
    &= \Pr(\bar{X} - p(1_a \cap 1_b) \geq \varepsilon) + \Pr(\bar{X} - p(1_a \cap 1_b) \leq - \varepsilon) \leq 2 \exp(-2n\varepsilon^2) \ .
\end{align}
Once inverted, this expression gives $n \leq \varepsilon^{-2}\log(2/\delta)/2$. Hence, estimating $f_{wW}(\I)$ from the rate of coincidences with uncertainty $\varepsilon$ and confidence level $1-\delta$, asymptotically requires $\mathcal{O}(\varepsilon^{-2}\log(2/\delta))$ statistical samples. The number of resources (photons) does not scale either with the input size $N$ or the number of hidden neurons $M$. For example, consider a bounded random variable $X$ with $M$ possible outcomes, each taking values in the interval $[0,1/2]$. As a worst-case scenario, let $X$ be uniformly distributed. With a bootstrap one can observe that, for a given $n$, the variance of the distribution of the sample average $\bar{X}$ is constant with respect to $M$. This means that the number of resources needed to estimate $\mathbb{E}(X)$ depends only on the desired accuracy and confidence level, not on the number $M$ of outcomes.

\section{CLASSICAL NEURAL NETWORK COUNTERPART}
\label{app:classical}
We discuss the classical counterpart of our quantum optical network, in two different scenarios. Our model imposes strict normalization constraints: each pure state must be normalized, while the mixture probabilities must be positive and sum to one. This corresponds to a $L^2$ normalization on the hidden parameters, and a positivity and $L^1$ constraints on the output weights, respectively. We benchmark the classical model both with and without such constraints.

In our simulations, we improved the classical model performances by fine-tuning the sigmoid activation function as $\tilde{\sigma}(x) = 1/[1+\exp(-11x+5.5)]$. \cref{fig:TrainingClassical} shows the results obtained for the classical model without constraints. The classification task remains the same as in Fig.~3: distinguishing between dog and plane images from the CIFAR-10 dataset. The results closely resemble those of the quantum optical network, with comparable accuracy and scaling with respect to number of hidden neurons. In \cref{tab:ClassicalComparison}a, we report the performance of the unconstrained classical model, while \ref{tab:ClassicalComparison}b shows the results in the constrained scenario. We note that the classical network can be successfully trained in both scenarios, although with different asymptotic accuracy.
\begin{table}[H]
	\centering
	\def\arraystretch{1.5}
	\setlength\tabcolsep{5pt}
        \begin{tabular}{|l|c|c|c|}
            \cline{2-4}
            \multicolumn{1}{c|}{\multirow{2}{*}{\parbox{1.5cm}{Accuracy\\(Best) [\%]}}} & \multirow{2}{*}{MNIST} & \multirow{2}{*}{\parbox{1.5cm}{\centering Fashion\\MNIST}} & \multirow{2}{*}{CIFAR} \\ \multicolumn{1}{c|}{} &&& \\ \hline
 	    M = 2  & 99.95 & 96.90 & 74.70 \\ \hline
 	      M = 16 & 99.95 & 96.80 & 78.45 \\ \hline
            M = 64 & 99.91 & 96.75 & 78.35  \\ \hline
           M = 128 & 99.95 & 96.80 & 77.65  \\ \hline
           M = 256 & 99.91 & 96.70 & 77.60 \\ \hline
          \multicolumn{4}{c}{\vspace{-0.25cm}} \\
          \multicolumn{4}{c}{(a)}
	\end{tabular}	
	\hspace{1cm}
	\begin{tabular}{|l|c|c|c|}
            \cline{2-4}
            \multicolumn{1}{c|}{\multirow{2}{*}{\parbox{1.5cm}{Accuracy\\(Best) [\%]}}} & \multirow{2}{*}{MNIST} & \multirow{2}{*}{\parbox{1.5cm}{\centering Fashion\\MNIST}} & \multirow{2}{*}{CIFAR} \\ \multicolumn{1}{c|}{} &&& \\ \hline
 	    M = 2  & 99.86 & 96.45 & 71.00 \\ \hline
 	      M = 16 & 99.86 & 96.55 & 74.50 \\ \hline
            M = 64 & 99.86 & 96.50 & 74.85  \\ \hline
           M = 128 & 99.91 & 96.50 & 75.05  \\ \hline
           M = 256 & 99.91 & 96.55 & 74.65 \\ \hline
          \multicolumn{4}{c}{\vspace{-0.25cm}} \\
          \multicolumn{4}{c}{(b)}
	\end{tabular}
\caption{\label{tab:ClassicalComparison}Best classification accuracy of a classical shallow network, for different numbers $M$ of hidden neurons, evaluated on the validation dataset. Training is performed with mini-batch gradient descent (without Adam) for $1000$ epochs, batch size $1000$ and learning rate $0.03$, on the datasets MNIST (zeros and ones), Fashion-MNIST (T-shirts/tops and pullovers) and CIFAR-10 (planes and dogs). (a) Unconstrained model. (b) Constrained model. Data simulated with TensorFlow \citep{tensorflow2015whitepaper}.}
\end{table}

\end{widetext}
\clearpage
\FloatBarrier
\bibliography{refs.bib}
\end{document}